# Updatable Queue Protocol based on TCP for Virtual Reality Environment

Ala'a Z. Al-Howaide[1], Mohammed I. Khaleel[2], Ayad M. Salhieh[3]

Jordan University of Science and Technology, Faculty of Computer and Information Technology, Irbid, Jordan
[1]computergy.alaa@gmail.com
[2]mohamed.i.k@live.com
[3]salhieh@just.edu.jo

## ABSTRACT

*The variance in number and types of tasks required to be implemented within Distributed Virtual Environments (DVE) highlighted the needs for communication protocols could achieve consistency. In addition, these applications had to handle an increasing number of participants and deal with the difficult problem of scalability. Moreover, the real-time requirements of these applications made the scalability problem more difficult to be solved. In this paper, we had implemented Updatable Queue Abstraction protocol (UQA) on TCP (TCP-UQA) and compared our implementation with the original TCP, UDP, and Updatable Queue Abstraction based on UDP (UDP-UQA) protocols. Results showed that TCP-UQA was the best in queue management.*

## 1. INTRODUCTION

Virtual Environments (VE) applications usually consisted of many different tasks that vary considerably from interfacing with input and output devices, through providing responsive user interaction, to simulating a dynamic environment. This variance in number and types of tasks required to be implemented within VE applications raised the idea of implementing VE applications over a distributed environments which in turn raised new tasks and issues to be considered such obtaining tracking and hand input information through continuous communication with external devices, rendering processes that required graphics intensive computation, and processes that performed collision detection between entities of the environment and simulated movement and behaviors for those entities. Distributed Virtual Environments (DVE) systems generally transmitted a great deal of state update messages between tasks and required only the most recent state available. In addition, old state information was not only useless in most cases; it was harmful for the running tasks and for the systems as a whole. So how information was communicated between tasks was critical.

DVE systems communicated through two types of communication modes. First, one-to-one communication, which was happened within processes that interface external devices, usually done using a client-server connections using the TCP/IP protocol, For example, Rubber Rocks application and the MR toolkit [1], which described a network software architecture for solving the problem of scaling very large distributed simulations. The fundamental idea was to logically partition virtual environments by associating spatial, temporal, and functionally related entity classes with network multicast group. The second was one-to-many communication, which happened among processes that shared a common state.

Within DVE there were three types of messages: the status update message, the command message, and the event message. The status update message represented a stream of state descriptions. The description may be, for example, a three dimensional position of the node, periodically and fast changing in the stock exchange, and so on.

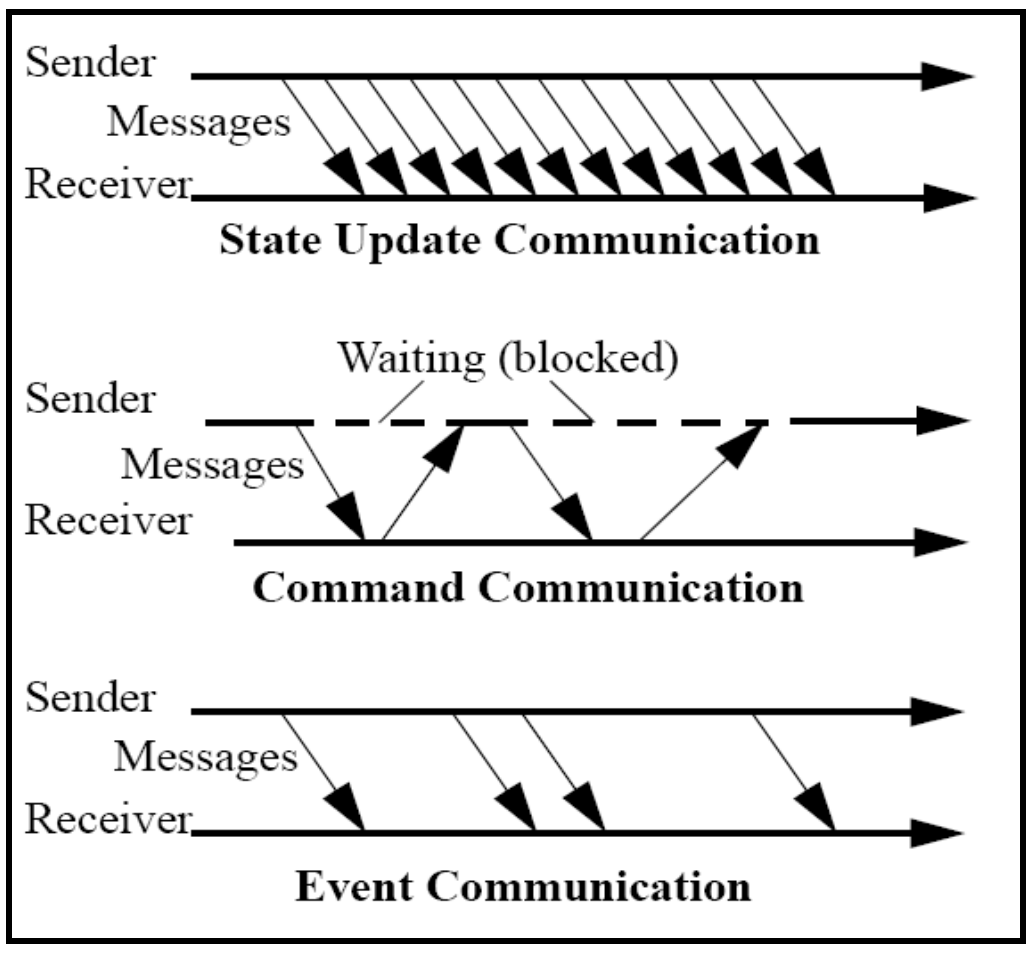

Figure 1. Messages types in DVE.

If the delivery of every state description message was guaranteed, then the receiving task would need to read all of the obsolete information before obtaining the latest state information (bandwidth cost), to decrease sending frequency one could implement a flaw control, but this would increase the average lag time (delay between sending and receiving a message). On the other hand, if messages delivery is not guaranteed, the same overflow problem occurred, but less frequently (whereas old messages would be automatically dropped if the messages were not received within a certain amount of time). [2]

The command message conveyed instructions from one task to another. Unlike the status message, the command message did not become obsolete. When the node sent a command message it must guarantee delivery to the First-In-First-Out (FIFO) order, so it required an acknowledgement message from the receiver. Furthermore, the command messages used to obtain state information from another task, thus, it used fewer messages than state update communication, but the client must block and wait for a response.[2]

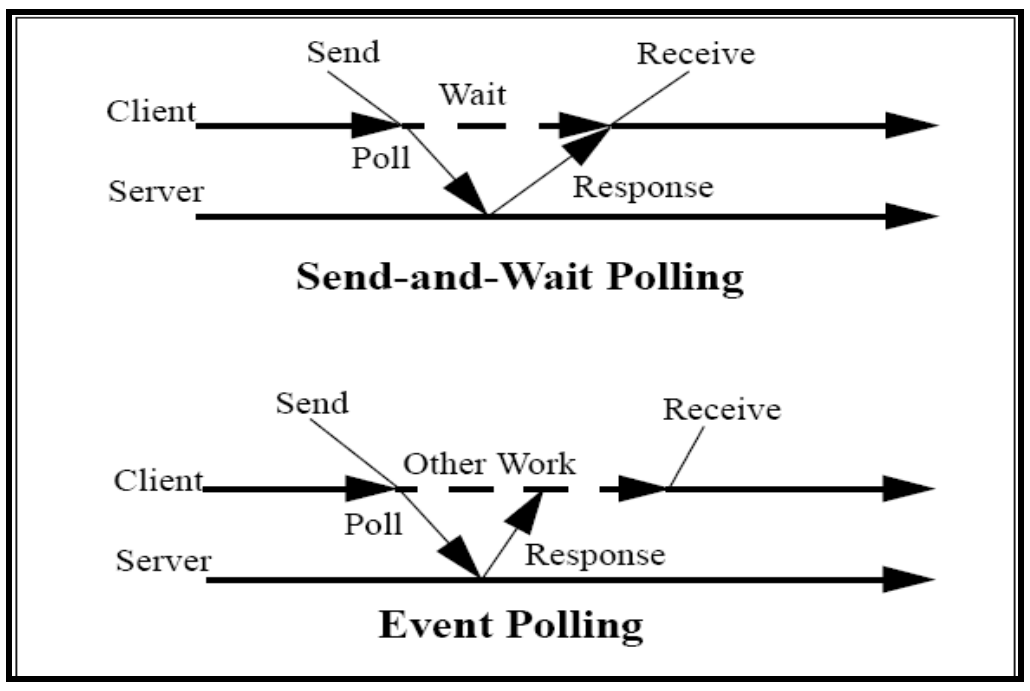

Figure 2. Command message.

The last message type was the event message which communicated the occurrence of events live key-presses; button presses, or object collisions. Unlike the command message, the event message did not require an acknowledgement message. As well as the command message, the event message was unique in that, it did not become obsolete and could occur frequently. The event message must be guaranteed and FIFO delivery order. In this type of messages, the client did not need to block for a response when it was used to obtain state information from another task, but if the response arrived early, then the information would become old. One could solve it by timing the arriving of the message which was difficult to predict. [2]

The rest of this paper was organized as follow: Section two discussed related work. Section three presented the implementation of the proposed model. Results were presented in section four. In section five, conclusion and the future work were presented.

## 2. RELATED WORK

The most common transport layer protocols that were used in distributed virtual environment communications were the Transport Control Protocol (TCP), which guaranteed reliable transmission across a distributed environment, and the User Datagram Protocol (UDP).

For short-term interactions, when the amount of information was small and it needed to be transmitted on an irregular basis, UDP was preferred over TCP. On the other hand, for long-term interactions, when data needed to be sent regularly or a large amount of data needed to be transmitted, then TCP was to be preferred. Furthermore, the UDP protocol did not guarantee message delivery to the receiving node. While the TCP, from the other hand, guaranteed the message delivery, but congested the network with the acknowledgement messages. Thus there was a trade-off between the quantity of data to be transmitted and the regularity of transmission. SPLINE [3], overcome the problem of the acknowledgements by sending messages with complete object state. Even if some messages were lost, consistency could be restored when the next message arrived.

Several protocols were used within distributed virtual environment trying to achieve consistency and message traffic issues. Status update message was one of the critical issues for distributed virtual environment due to its importance and impact. The status update message was the only way that defined and kept track the recent status of each component in the distributed virtual environment. Without having the recent state, the tasks could not be accomplished at all or it would be not correct where resources spent on performing on obsolete information. The TCP/IP protocol was used in the DVE but it did not suit all types of these applications because it did not allow messages to be deliberately dropped. This forced any task requested the recent state of another task to hear all of its obsolete states, which would be time and bandwidth consuming. On the other hand, it was necessary to use TCP with applications required very high degree of objects stat accuracy, such as virtual medical environments. UDP allowed dropping messages which decreased the message traffic, but it did not allow the application to decide which messages to be dropped. Distributed Interactive Simulation (DIS) protocol which was used by NPSNET [4], which was developed on a peer-to-peer network topology, used UDP Broadcast or multicast communication. It simulated other peers through using the *dead reckoning* algorithm, which simulated state changes locally so a peer would only broadcast its state when it had changed within a threshold error from what other processes would simulate. This protocol was not a general solution to communication between distributed virtual environment tasks because event and command communication required ordered and guaranteed message passing. In addition, the receiver should not had to wait if the first send was unsuccessful, also it might at finishing transmitting, the state became inconsistent if it stopped transmitting at specific point. Finally, it allowed obsolete messages/states to be resided in the receiver's message buffer. A protocol which was proposed by Holbrook [4], a variation of DIS, allowed the *sender* to decide which messages needed to be re-sent at any point in time, and did not allow old messages to be buffered at the receiver where they could become obsolete through using unguaranteed message passing, allowing the *receiver* to decide if a lost message, once it was detected, needs to be resent. It used "heartbeat" messages to ensure message loss detection, and message log processes to assist in message recovery.

A Stream Control Transmission Protocol (SCTP) [5], was a new protocol similar to TCP as it provides reliable full-duplex connection, in other words, it was a simultaneous two-way data or voice transmission. In addition, it offered new delivery options that suit multimedia applications such as multi-streaming. Also, it allowed independent delivery among data streams, multi-homing, allowing end points of a single association to have multiple IP addresses and *partial reliability* and it allowed each message to be assigned a delivery reliability level.

G. Drew [6], proposed a network communication protocol for distributed virtual environment which called an Updatable Queue Abstraction (UQA). The UQA was a simple First In First Out (FIFO) queue with the ability of storing additional information, called keyed data, attached with each message. These data gave each message a special key to represent its type (command, event, of status), and to represent the process that sent the message (process A, process B, etc). The main idea of the UQA was when a node received the status message; the message was replaced with the latest status message of the same sender that stored in its queue. They tested the model in one-to-one communication fashion.

The network layer protocol that provided simultaneous unreliable transmission to multiple destination nodes was called multicast protocols. The multicast protocols were used whenever the messages needed to be sent to multiple nodes. Within DVE, multicast protocols were useded extensively as a filtering process.

One of the popular multicast protocols was the overlay multicast. The overlay multicast protocol was implemented at the application layer. It was preferred to the Overlay multicast protocol used over the networks that did not offer multicast capability at the network layer. The best known overlay network protocols were the Multicast Backbone (MBone) [7] and the DIVEBone [8] which extended the MBone as a stand-alone part of the DIVE toolkit.

Another multicast protocol called reliable multicast. Reliable multicast protocol referred to error-free eventual delivery of information to all the application with same level of ordering. In reliable multicast each member of the multicast group was responsible for its own correct reception of all the data using a time-out mechanism.

Sato [9] proposed a reliable multicast protocol suitable for DIAs. The proposed protocol used the concept of Mutual Aid Regions (MAR). Each MAR had a node that was responsible for acknowledging receipt of packets in that MAR and that sent timeout requests to the nearest MAR if a packet was not received.

Another approach of multicast protocol was the Scalable Communication Protocol for Large-Scale Virtual Environments (SCORE) [10]. SCORE was a scalable multicast-based communication protocol for Large-Scale Virtual Environments (LSVE) work on the Internet. This approach involved the dynamic partitioning of the virtual environment into spatial areas and the association of these areas with multicast groups. It used a method based on the theory of planar point processes to determine an appropriate cell-size, so that the incoming traffic at the receiver side remained with a given probability below a sufficiently low threshold.

Multicast protocols had some challenges. First, due to issues such as addressing, many internet routers were limited in the number of groups they can support or they were not multicast aware. Second, it was difficult to implement efficiently on a point to point medium. Finally, it was costly to be controlled and to be administrated the congestion of the network that results from flooding the network by multicasting. [11]

## 3. IMPLEMENTATION

QualNet network simulator was used in our implementation for the distributed virtual environment system. QualNet was a commercial network simulation tool implemented in C++ that simulated wireless and wired packet mode communication networks. QualNet was used in the simulation of MANET, WiMAX networks, satellite networks, wireless sensor networks, etc.

It had a graphical user interface and a set of library functions used for network communication. [13]

The implementation involved three phases. The first phase was generating three types of messages. Since the only message that became obsolete and needed to be replaced was the status update message, wherefore, we did not need to distinguish between the other two types which were the command and the event message. So, we implemented two types of messages that could be sent by the traffic generator node which were the status update messages as one type, and the command or event messages as another type.

Phase two involved the implementation of the traffic generator. A traffic generator was the node that generated and sent the messages to other nodes in the distributed virtual environment. We used probability theory to control the traffic generation of the messages. Based on our knowledge, the status messages represented about two thirds of the total number of messages being sent between two nodes. So we implemented the probability of sending the status update message as 70%, while the probability of sending command or event message is 30%. We used the random function to generate the number that used as a probability. The reason in using the random number probability generator, rather than uniform message generator, was to insure the realistic of two task sending information in the real world.

The third phase was implementing the updatable queue abstraction. The updatable queue abstraction algorithm was shown in figure 3 and discussed as follows.

When the node received the message, it first tested if the message was command/event message. If so, then it would insert it into the tail of the queue and increase the tail of the queue by one. Otherwise, the message was a status message. Then it would retrieve the last message inserted into the queue and check if it was a command or event message. If so, then the node would insert the retrieved message followed by the new message into the queue. Otherwise, the node should check if the retrieved message had the same ID number as the new message, in other words, if the message at the end of the queue and the new messages were from the same sender. If the messages were from different senders then the node would insert the retrieved message and the new message into the queue. Otherwise, if both status messages from the same sender, the node would remove the retrieved messages and considered it as an obsolete message and insert the new status message into the queue.

```
If the message is command or event message
        Then insert the message to the queue;
Else
        Retrieve the last message from the queue using
        "retrieve" function;
        If the retrieved message is command or event
        message
                Then insert the retrieved message and the
                new message to the queue;
        Else
                If the retrieved message from different
                sender
                        Then insert the retrieved message and
                        the new message to the queue;
                Else
                        Remove the retrieved message and
                        insert only the new message to the
                        queue;
```

Figure 3. Updatable queue abstraction algorithm.

We applied our implementation on both transport layer communication protocols which were TCP and UDP and called the new modified protocols TCP_UQA and UDP_UQA respectively. Our implementation tested with both one-to-one communication and one-to-many communication as shown in figures 4 and 5 respectively.

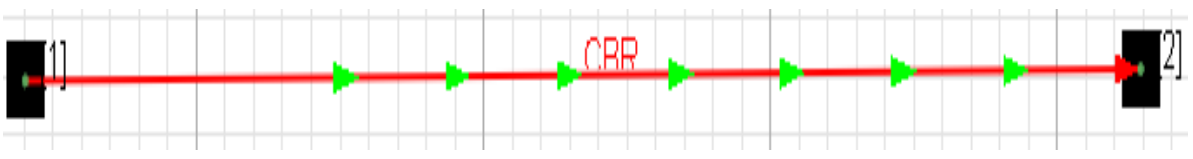

Figure 4. One-to-one communication protocol.

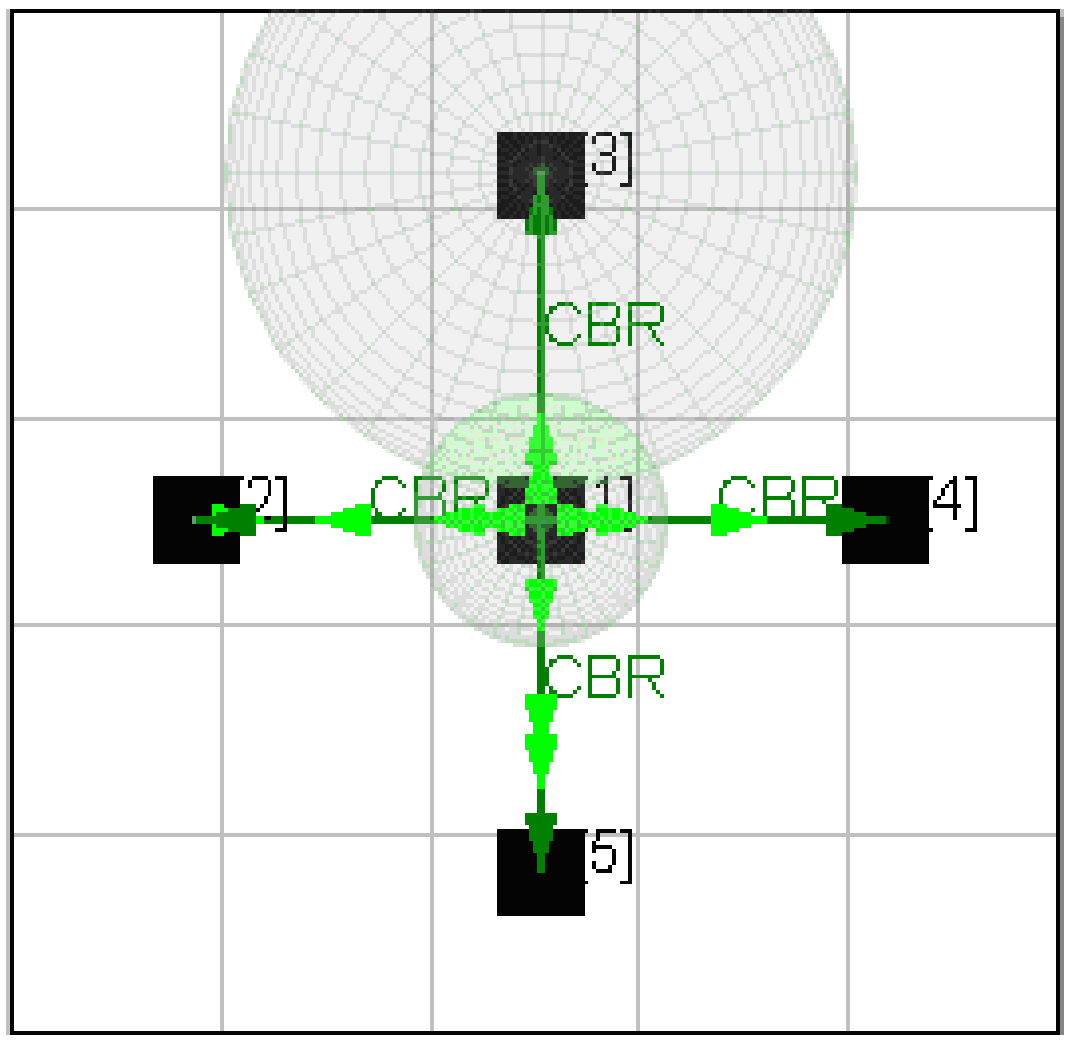

Figure 5. One-to-many communication protocol.

## 4. EXPERIMENTS

To measure the efficiency of VE protocols we perform 96 experiments with the TCP, UDP, TCP-UQA, and UDP-UQA on different packet sizes (32, 256, 512 bytes), where four values (0.0, 0.033, 0.05, 0.1 second) of delay at the receiver were used to represent processes which spend different amounts of time working on other problems than communication.

We have conducted two sets of experiments. The first set was carried to test one-to-one communication between two nodes as in figure 4. Each protocol was tested on the three different packet sizes and on the four different receiver delay values. The source sent 1000 messages to the destination with a 70% probability the message would be a status message and 30% probability to be command and event. Each experiment was run for 180 seconds in which all the 1000 messages were sent by the source.

The second set was carried to test one-to-many communication between one source and four destinations as in figure 5. All of the destination nodes were on the same distance from the source node. Each protocol was tested on the three different packet sizes and on the four different receiver delay values. The source sent 1000 messages for each destination with a 70% probability the message would be a status message and 30% probability to be command and event. Each experiment was run for 720 seconds in which all the 4000 messages were sent by the source.

## 5. RESULTS

To study the efficiency of the tested protocols we measured the average value of all experiments carried on the three different packet sizes (32, 256, 512 bytes). Figures 6 to 10 show the one-to-one communication results and Figures 11 to 13 show the one-to-many communication results.

For one-to-one communication, Figure 6 presented the results of average client throughput. The UDP and UDP-UQA protocols showed the best results for all packet sizes at the all receiver delays, because the client would only require transmitting the requested messages without any additional packets. While TCP protocol spent additional time in transmitting acknowledgements for each packet it sends. TCP-UQA showed the worst results for all cases, because in addition to the time spent in transmitting acknowledgements it spent additional time in updating the queue.

Figure 7 presented the average server throughput. TCP protocol was the worst because of sending acknowledgements, so the server is forced to process the acknowledgements in addition to the messages. While TCP-UQA protocol showed better results, because of sending less messages and thus fewer acknowledgements. Both UDP and UDP-UQA protocols were the best, Both nearly showed the same results in all test cases, except that UDP-UQA protocol was not efficient when sending packets of size 256 bytes. Unless of this impact the average server throughput of UDP-UQA still better than TCP and TCP-UQA protocols.

From figures 8, 9, 10, a closer look can be carried on the queue management of each protocol. UDP had the longest average queue and the highest difference from the average peak queue size, which meant that the queue size was not stable and all messages arrived at once to the queue which in turn increased the average time of the message in the queue. While UDP-UQA protocol showed better results only because of dealing with fewer number of messages but still suffer from unstable queue size and low queue service rate. Surprisingly, Both TCP and TCP-UQA protocols showed shorter average queue lengths and smaller peaks than both UDP and UDP-UQA protocols because of staggering [12]. TCP-UQA was the most stable queue and the highest queue service rate.

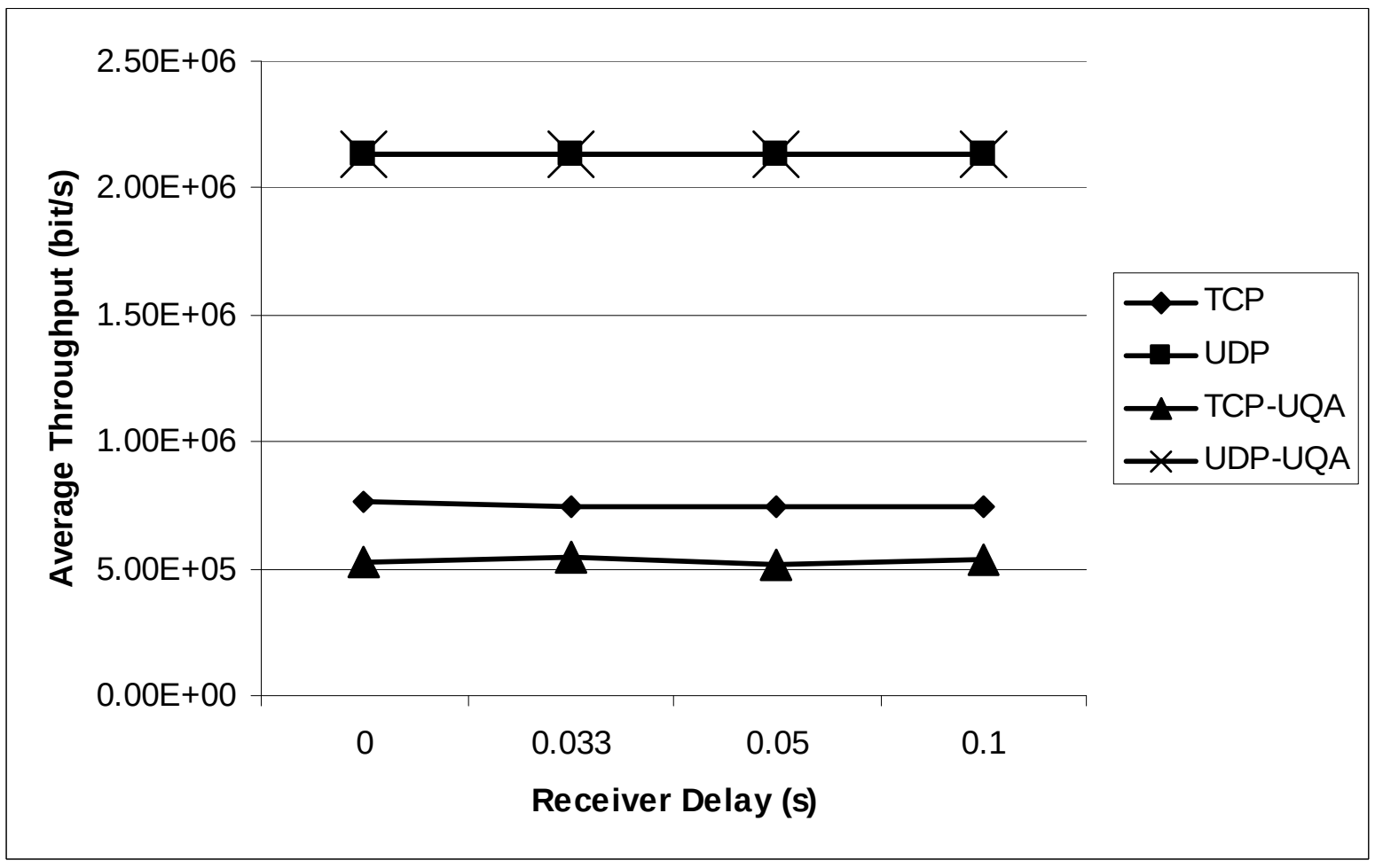

Figure 6: One-To-One average client throughput (bit/s).

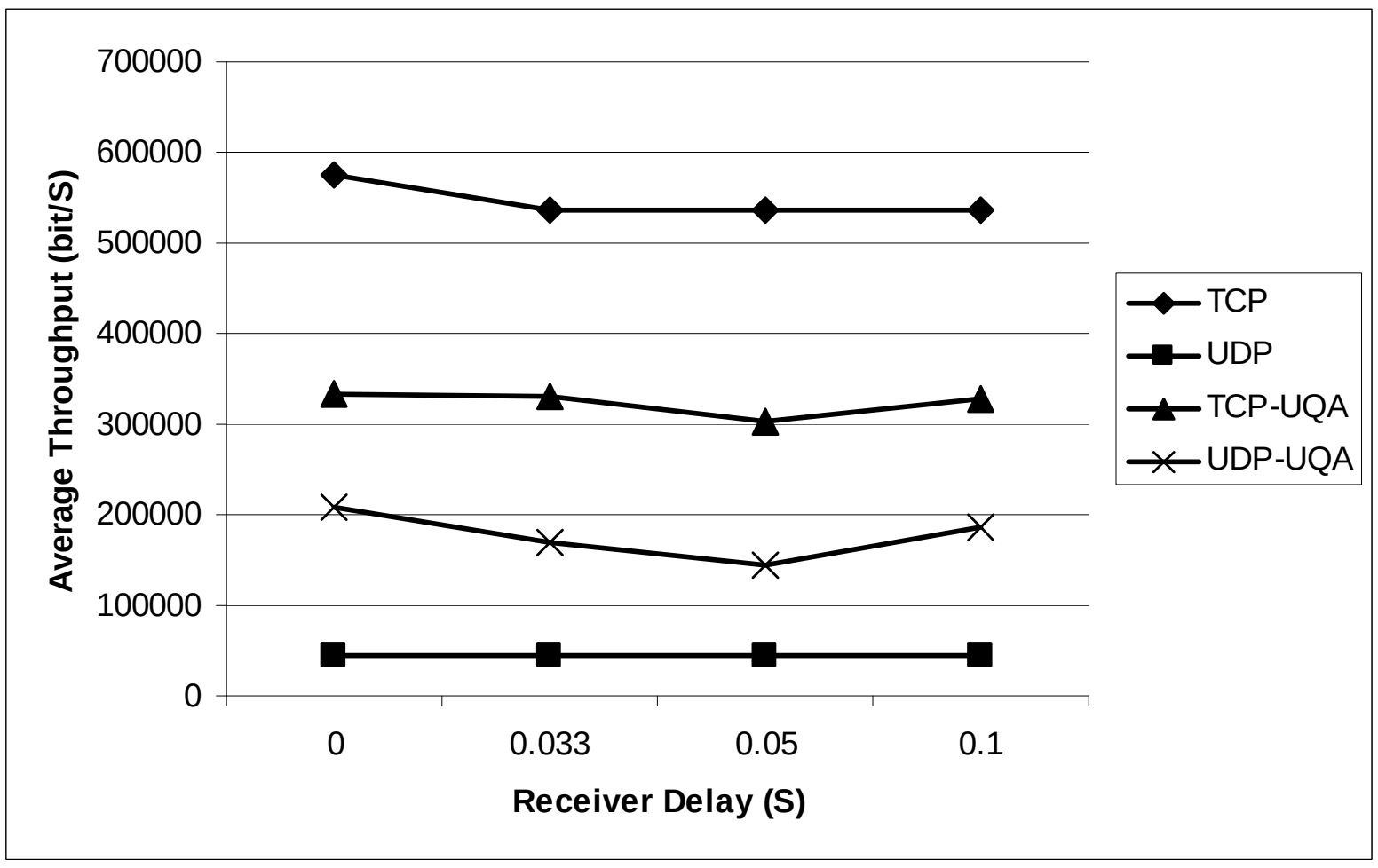

Figure 7: One-To-One average server throughput (bit/s).

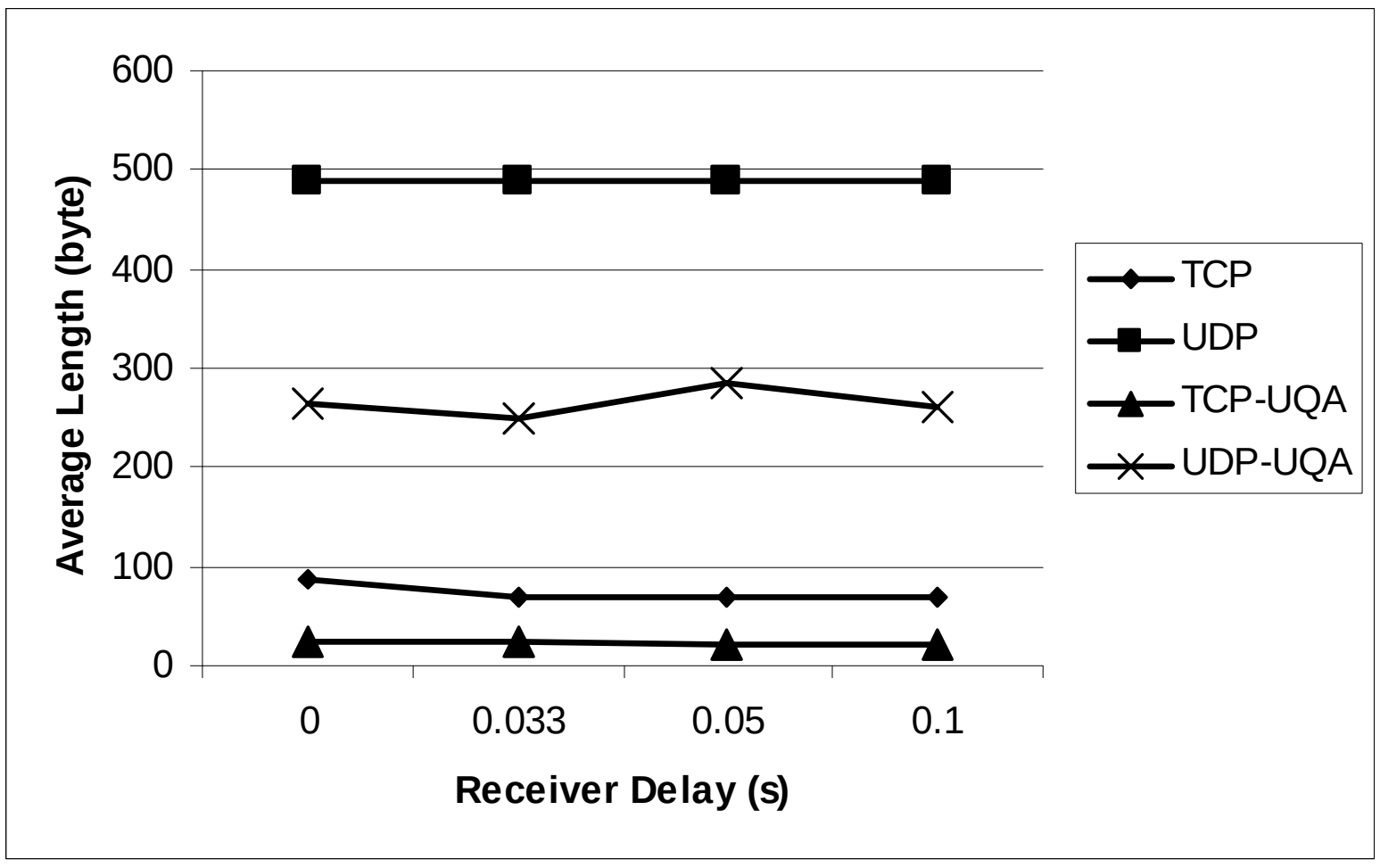

Figure 8: One-To-One average queue length.

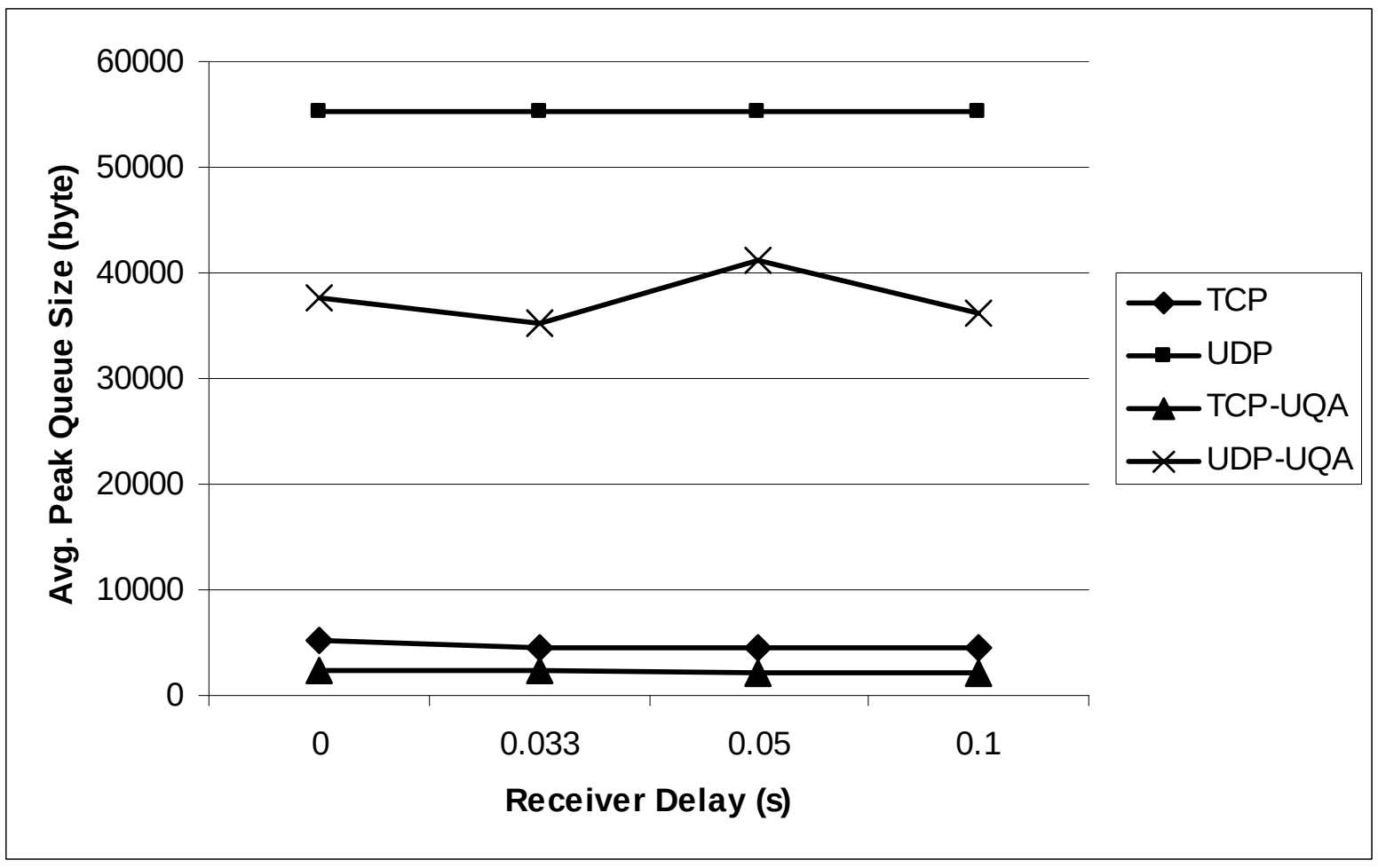

Figure 9: One-To-One average peak queue size.

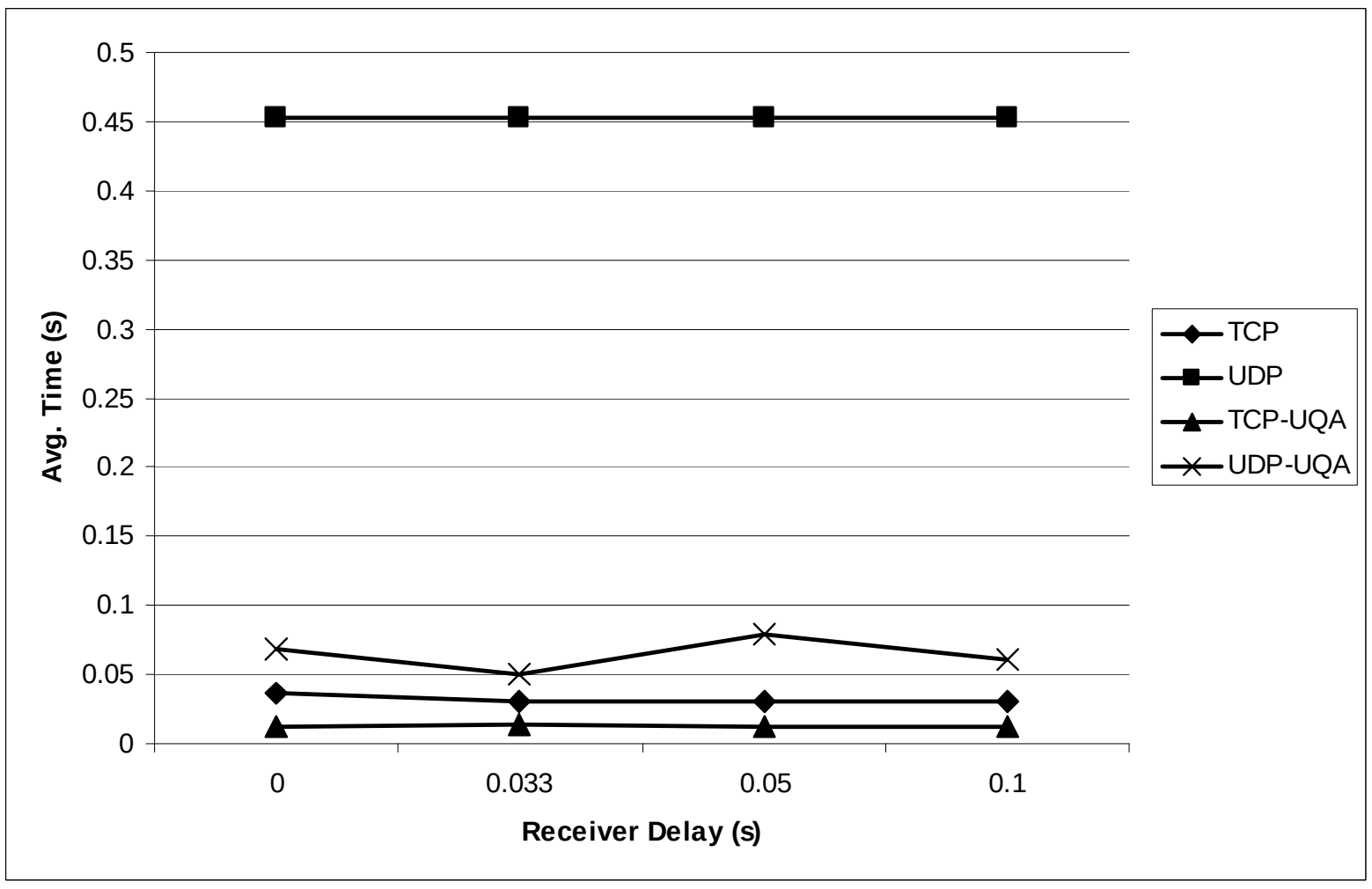

Figure 10: One-To-One average time in the queue.

For one-to-many communication, figure 11 presented the results of average queue length. The best protocol was TCP-UQA for all test cases. While UDP protocol was the worst at the receiver delay value of 0.033, 0.05, and 0.1 second. UDP-UQA protocol showed results better than both UDP and TCP protocols. From figures 12 and 13 we could see that UDP-UQA showed more stability than one-to-one communication but still suffer from slow queue service rate because of congestion of messages in the queue. UDP-UQA protocol showed better results than UDP protocol in all test cases.

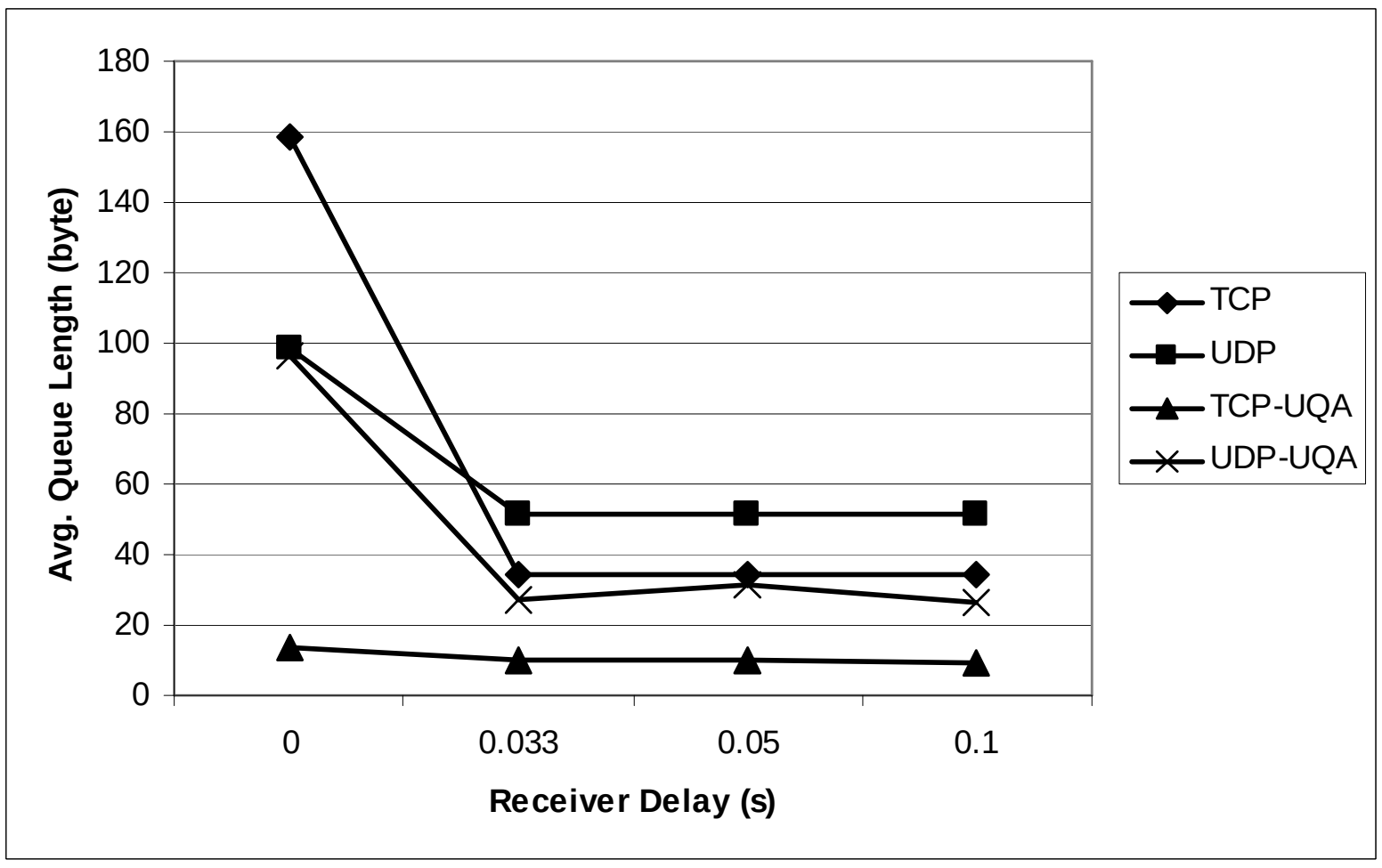

Figure 11: One-To-Many average queue length.

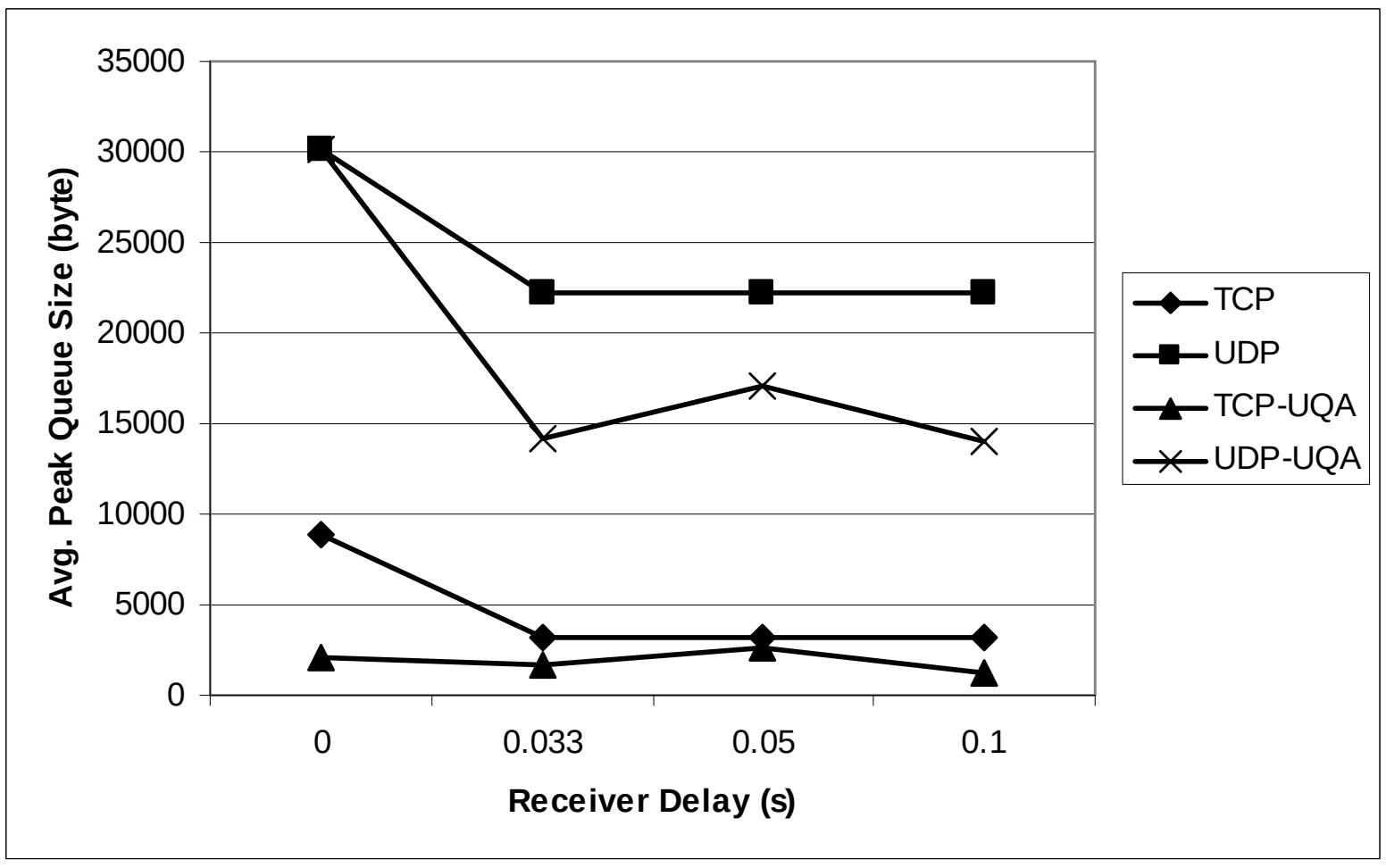

Figure 12: One-To-Many average peak queue size.

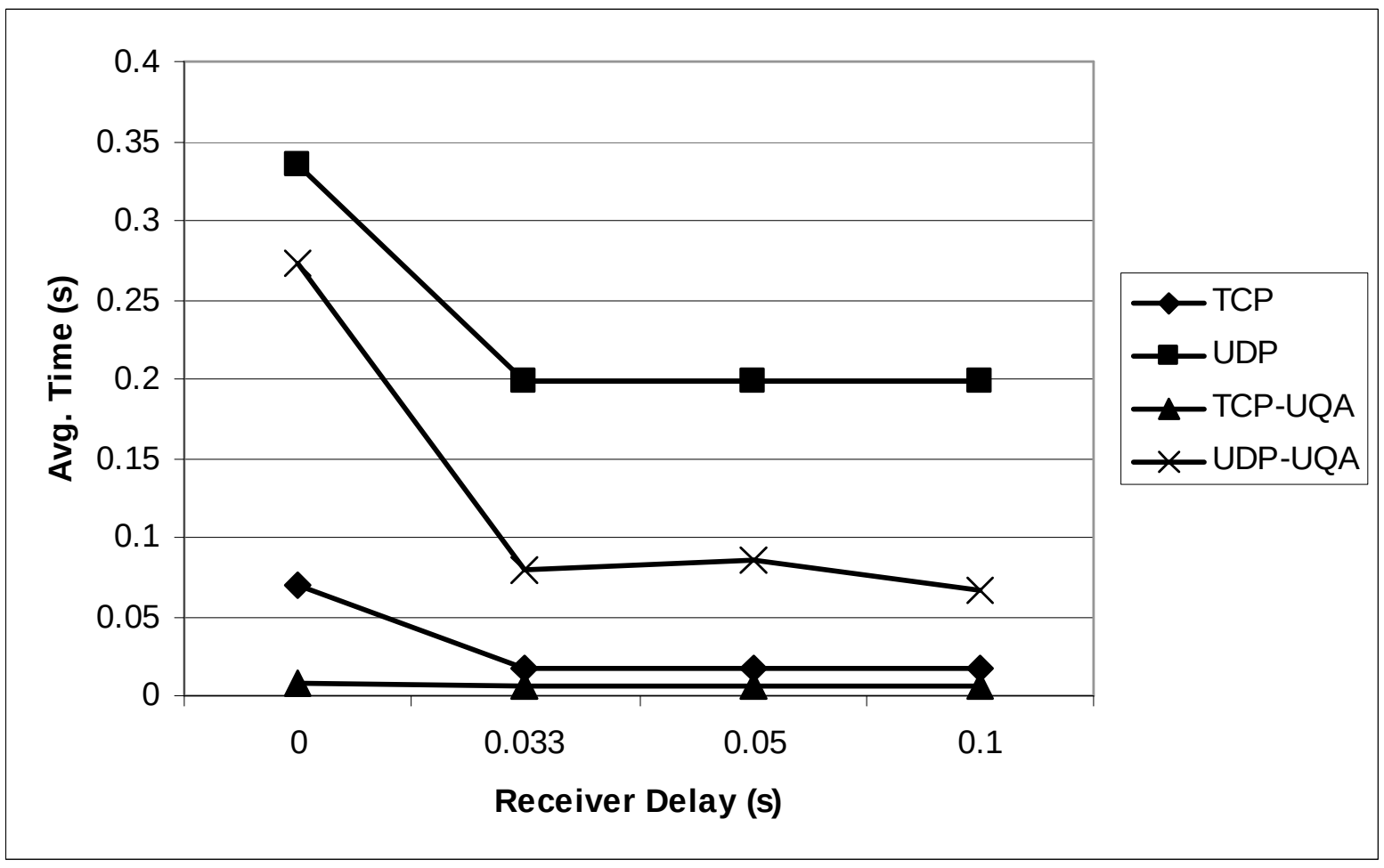

Figure 13: One-To-Many average time in the queue.

## 6. CONCLUSION

Distributed virtual environments raise new issues to the communication protocols upon the variety of tasks and variety of messages. Several protocols were proposed to improve and meet DVE communication requirements.

In this paper we have implemented UQA on both the original TCP and UDP protocols. Surprisingly, TCP-UQA showed better queue management than other protocols for both small and large messages. This advantage of TCP-UQA makes it a good candidate when each object state is important and cannot be neglected or missed such considering virtual medicine environments, while UDP-UQA did not show a tangible difference in its performance when compared with UDP. TCP-UQA was slower than other protocols in some cases because of its additional processing on each message; this disadvantage makes it a bad candidate for time-restricted application.

All test cases were run on uni-cast protocols. UQA have many advantages can benefit from by implementing it on multicast protocols. We will consider this for future work.

## 7. References

[1] Macedonia M. R., M. J. Zyda, D. R. Pratt, D. P. Brutzman, and P. T. Barham, (1995), "Exploiting Reality with Multicast Groups: A Network Architecture for Large-scale Virtual Environments", *Proceedings of Virtual Reality Annual International Symposium (VRAIS '95), Proc. of IEEE*, pp. 2-10.

[2] Andrew S. Tanenbaum, Maarten Van Steen, (2006), "Distributed Systems: principles and Paradigms", Second Edition.

[3] Barrus, J. W., Waters, R. C. & Anderson, (1996), "Locales and Beacons: Efficient and Precise Support for Large Multi-user Virtual Environments", *Proceedings of Virtual Reality Annual International Symposium (VRAIS '96). IEEE, Santa Clara, CA*, pp. 204-212.

[4] Holbrook, Hugh W., Sandeep K. Singhal, and David R. Cheriton, (1995), "Log-Based Receiver-Reliable Multicast for Distributed Interactive Simulation", *Proc. of ACM SIGCOMM (Cambridge, MA, Aug. 1995)*, pp. 328-341.

[5] Caro, A., L. Jr, Iyengar, J. R., Amer, P. D., Ladha, S., Heinz, G. J. I. & Shah, K. C., (2003), "SCTP: A proposed Standard for Robust Internet Data Transport*", IEEE Computer*, pp. 56-63.

[6] Kessler G. D., Larry F. Hodges, (1996) "A Network Communication Protocol for Distributed Virtual Environment Systems", *Proceedings of Virtual Reality Annual International Symposium (VRAIS '96). IEEE, Santa Clara, CA*, pp. 214-221.

[7] Eriksson, H. (1994). "MBone: The Multicast BackBone". *Communications of the ACM*, pp. 54-60.

[8] Frécon, E. (2003). "DIVE: A generic tool for the development of virtual environments", *Proceedings of 7th International Conference on Telecommunications (ConTEL 2003)*, pp. 345-352.

[9] Sato, F., Minamihata, K., Fukuoka, H. & Mizuno, T., (1999), "A reliable multicast framework for Distributed Virtual Reality Environment", *Proceedings of International Workshop on Parallel Processing*, Aizu-Wakamatsu, Japan,, pp. 588-593

[10] E. Lety, T. Turletti, and F. Baccelli, (2004), "SCORE: A Scalable Communication Protocol for Large-Scale Virtual Environments", *IEEE/ACM Trans. Net*, pp. 247–60

[11] Declan Delaney, Tomas Ward, and Seamus McLoone, (2006) "On Consistency and Network Latency in Distributed Interactive Applications: A Survey- Part II", *Teleoper Virtual Environ.*, pp. 465– 482.

[12] Alison Carrington, Chris Harding, and Hongnian Yu, (2008), "Optimising Wireless Network Control System Traffic – Using Queuing Theory", *Proceedings of the 14th International Conference on Automation & Computing, Brunel University*, West London, UK,.

[13] http://www.scalable-networks.com, 2010.